\documentclass[twocolumn,english,pre]{revtex4-1}
\usepackage[T1]{fontenc}
\usepackage[latin9]{inputenc}
\usepackage{amssymb}
\usepackage{graphicx}
\usepackage{esint}
\usepackage{times}

\makeatletter

\@ifundefined{textcolor}{}
{%
 \definecolor{BLACK}{gray}{0}
 \definecolor{WHITE}{gray}{1}
 \definecolor{RED}{rgb}{1,0,0}
 \definecolor{GREEN}{rgb}{0,1,0}
 \definecolor{BLUE}{rgb}{0,0,1}
 \definecolor{CYAN}{cmyk}{1,0,0,0}
 \definecolor{MAGENTA}{cmyk}{0,1,0,0}
 \definecolor{YELLOW}{cmyk}{0,0,1,0}
 }

\makeatother

\usepackage{babel}
\begin{document}

\title{Nonlinear elastic model for faceting of vesicles with soft grain
boundaries}

\author{Rastko Sknepnek}
\email{sknepnek@gmail.com}
\affiliation{Department of Materials Science and Engineering, Northwestern University, Evanston, Illinois 60208}
\author{Monica Olvera de la Cruz}
\email{m-olvera@northwestern.edu}
\affiliation{Department of Materials Science and Engineering, Northwestern University, Evanston, Illinois 60208}
\selectlanguage{english}

\begin{abstract}
We use an elastic model to explore faceting of solid-wall vesicles
with elastic heterogeneities. We show that faceting occurs in regions
where the vesicle wall is softer, such as areas of reduced wall thicknesses
or concentrated in crystalline defects. The elastic heterogeneities
are modeled as a second component with reduced elastic parameters.
Using simulated annealing Monte Carlo simulations we obtain the vesicle
shape by optimizing the distributions of facets and boundaries. Our
model allows us to reduce the effects of the residual stress generated
by crystalline defects, and reveals a robust faceting mechanism into
polyhedra other than the icosahedron.
\end{abstract}
\maketitle
Nature uses various vesicular structures to compartmentalize and differentiate
matter in order to achieve a desired set of functions. Depending on
their origin and function, vesicles can take a wide variety of shapes
and sizes with very different physical and chemical properties. A
prototypical example of a faceted structure is the capsid of large
viruses, for which an icosahedral shape was elegantly described by
Caspar and Klug \cite{Caspar62}. Next to these regular polyhedra,
bacterial microcomponents such as carboxysomes \cite{Iancu2010} (found
in cyanobacteria \cite{yeates2008protein}, for example), are known
to take on irregular faceted shapes. X-ray diffraction of small ($\sim30$nm
diameter) single-walled DPPC vesicles found \cite{blaurock1979small}
that below the thermal transition they facet into polyhedra. Thick-wall
nested fullerenes are also known to sometimes take faceted shapes
\cite{srolovitz1995morphology}. Faceted shapes were recently observed
\cite{Greenfield2009} by coassembling oppositely charged amphiphiles.
It has been argued that under the right conditions electrostatic correlations
can lead to faceting \cite{Vernizzi07a} or even a complete collapse
of the structure \cite{Sknepnek2011}.

The existence of facets separated by sharp edges that have been observed
in experiments can be understood in terms of the vesicle wall being,
at least in part, solid \cite{blaurock1979small} or as a result of
excess amphiphiles exhibiting spontaneous curvature being present
in sufficient quantity \cite{Haselwandter2010,Haselwandter2011}.
Molecules within a solid wall have fixed connectivity and the vesicle
can sustain shear, bending, and stretching deformations. The microscopic
structure and properties of the ordered phase is intimately related
to the vesicle topology. Very interesting phenomena occur if the local
molecular order favored by interactions is not compatible with the
global topology and thus cannot be realized simultaneously throughout
the whole surface \cite{bowick2009two}. For example, the preferred
planar ordering of identical particles interacting with a spherically
symmetric pair potential is a hexagonal lattice, where each site has
exactly six equidistant nearest neighbors. In other words, it is possible
to tile a plane with equilateral triangles. However, this is not the
case for a sphere, and topological defects (i.e., sites with coordination
$z\neq6$) are necessarily present. For a triangular tessellation
of a sphere, Euler's theorem ensures that there are at least 12 sites
that are five-coordinated \cite{bowick2009two}. These disclination
defects repel each other and take a conformation that maximizes their
mutual separation, thus positioning themselves at the vertices of
an inscribed icosahedron. 

The stretching energy of a five-fold disclination on a plate of radius
$r$ is $\propto r^{2}$ \cite{Seung88}. If the plate is allowed
to buckle out of plane it can reduce the stretching energy at the
expense of a bending penalty that is $\propto\log r$ \cite{Seung88}
and form a cone with the apex centered at the defect. On a sphere
these two energies follow the same scaling laws, albeit with different
prefactors \cite{Bowick2000}. If the sphere is sufficiently large
it can lower its energy by buckling into an icosahedron via a similar
conical deformation seeded at the twelve defects \cite{Lidmar03}.
Recently we have shown that if a vesicle is assembled of multiple
molecular species with different elastic properties, depending on
the relative ratio and the line tension of the components, it can
take a number of regular and irregular polyhedral \cite{Vernizzi2011}
or Janus-like shapes \cite{sknepnek2011buckling}. 

These results suggest that a more general buckling mechanism is possible
when the vesicles are not elastically homogeneous. However, in a previous
work \cite{Vernizzi2011} we studied two-component elastic shells
with the twelve five-fold defects fixed to the corners of an inscribed
icosahedron. Here we show that faceting occurs even if such restrictions
are lifted. In this paper we introduce a nonlinear elastic model to
study faceting of a vesicle with solid domains connected with soft
boundaries. We argue that material heterogeneities, such as local
variations in the wall thickness or due to defects in the crystalline
order, can lead to local softening of the vesicle wall and provide
a pathway for lowering the energy. In order to satisfy both, the preferred
local flatness of the solid domains and the curved global geometry
the curvature is focused along the softer domain boundaries, which
leads to faceting of the entire vesicle. 

We assume that the vesicle wall is thin compared to its radius, $R$,
and can be represented as its two-dimensional midsurface. Such an
approximation is indeed justified as the vesicle wall is typically
a few nanometers thick compared to the radius that ranges from tens
of nanometers to microns. The midsurface can be parametrized with
two parameters $s^{1}$ and $s^{2}$, as a locus of points $\mathbf{r}=\mathbf{r}\left(s^{1},s^{2}\right)$.
The metric tensor is defined as $g_{\alpha\beta}=\partial_{\alpha}\mathbf{r}\cdot\partial_{\beta}\mathbf{r}$,
where $\alpha,\beta=1,2$. The second fundamental form, related to
the surface curvature, is given by $b_{\alpha\beta}=\partial_{\alpha\beta}\mathbf{r}\cdot\mathbf{n}$,
with $\mathbf{n}$ being the unit normal. The elastic energy of the
midsurface can be written as \cite{koiter1966nonlinear}
\begin{equation}
E=\int dA\mathcal{A}^{\alpha\beta\gamma\delta}\left(\frac{h}{2}u_{\alpha\beta}u_{\gamma\delta}+\frac{h^{3}}{24}b_{\alpha\beta}b_{\gamma\delta}\right),\label{eq:elstic_energy}
\end{equation}
where $u_{\alpha\beta}=\frac{1}{2}\left(g_{\alpha\beta}-\overline{g}_{\alpha\beta}\right)$
is the strain tensor, $\overline{g}_{\alpha\beta}$ is a reference
metric, $dA=\sqrt{\left|g\right|}ds^{1}ds^{2}$ is the area element,$\left|\cdot\right|$
is the determinant, $h\ll R$ is the vesicle thickness and the summation
over pairs of repeated indices is assumed. Compared to a recent study
\cite{dias2011programmed}, here we follow Koiter's arguments \cite{koiter1959consistent}
and retain only terms consistent with the Kirchhoff-Love assumptions
\cite{efrati2009elastic} of a negligible normal stress. The rank-four
elastic tensor $\mathcal{A}^{\alpha\beta\gamma\delta}$ is determined
by the material properties. If the material is isotropic $\mathcal{A}^{\alpha\beta\gamma\delta}$
depends only the Young's modulus $Y$ and Poisson's ratio $\nu$,
and $\mathcal{A}^{\alpha\beta\gamma\delta}=\frac{Y}{1+\nu}\left(\frac{\nu}{1-\nu}\overline{g}^{\alpha\beta}\overline{g}^{\gamma\delta}+\overline{g}^{\alpha\gamma}\overline{g}^{\beta\delta}\right)$,
with $\overline{g}^{\alpha\gamma}\overline{g}_{\gamma\beta}=\delta_{\beta}^{\alpha}$
and $\delta_{\beta}^{\alpha}$ being the Kronecker delta symbol.

The first term in Eq.~(\ref{eq:elstic_energy}) represents energy
associated with stretching deformations while the second term accounts
for bending. For an isotropic material the stretching and bending
energies simplify to 
\begin{eqnarray}
E_{s} & = & \frac{h}{2}\int dA\frac{Y}{1+\nu}\left(\frac{\nu}{1-\nu}u_{\alpha}^{\alpha}u_{\beta}^{\beta}+u_{\alpha}^{\beta}u_{\beta}^{\alpha}\right),\label{eq:stretch}\\
E_{b} & = & \frac{h^{3}}{24}\int dA\frac{Y}{1+\nu}\left(\frac{\nu}{1-\nu}b_{\alpha}^{\alpha}b_{\beta}^{\beta}+b_{\alpha}^{\beta}b_{\beta}^{\alpha}\right),\label{eq:bend}
\end{eqnarray}
with $u_{\alpha}^{\beta}=\overline{g}^{\beta\gamma}u_{\alpha\gamma}$
and $b_{\alpha}^{\beta}=\overline{g}^{\beta\gamma}b_{\alpha\gamma}$.
With the mean curvature $H\equiv\frac{1}{2}b_{\alpha}^{\alpha}$ and
the Gaussian curvature $K\equiv\det\left(b_{\alpha}^{\beta}\right)$,
the bending energy can be rewritten as $E_{b}=\int dA\kappa\left(2H^{2}-\left(1-\nu\right)K\right)$,
where $\kappa=\frac{h^{3}}{12}\frac{Y}{\left(1-\nu^{2}\right)}$ is
the bending rigidity. The prefactor of the Gaussian curvature term
is negative and proportional to $\kappa$, thus suppressing saddle-like
($K<0$) conformations. For a homogeneous vesicle with fixed topology
the Gauss-Bonnet theorem ensures that $\int dAK=\mathrm{const.}$
and the Gaussian curvature term can be omitted. However, for a multicomponent
vesicle, as is the case here, this term must be retained. 

Variation of Eq.~(\ref{eq:elstic_energy}) with respect to $u_{\alpha\beta}$
and $b_{\alpha\beta}$ leads to a set of nonlinear partial differential
equations for the mechanical equilibrium \cite{efrati2009elastic,dias2011programmed}.
Instead of directly solving those equations, which is a formidable
task even in simple geometries, we opt for a numerical minimization
of the energy, Eq.~(\ref{eq:elstic_energy}), via Monte Carlo (MC)
simulations. The vesicle is represented as a discrete triangular mesh
with two types of triangles, \emph{soft} and \emph{hard}. From Eqs.~(\ref{eq:stretch}) and (\ref{eq:bend})
it is evident that for a
fixed Young's modulus $Y$ and Poisson's ratio $\nu$ both bending
and stretching moduli depend only on the vesicle wall thickness, $h$.
Therefore, for simplicity, we assume that the elastic properties of
a component are determined by its thickness, with the soft component
being thinner. This allows us to explore the phase behavior as a function
of only two parameters, relative thickness $\eta=\frac{h_{hard}}{h_{soft}}$
and the fraction $f$ of the soft component. We keep in mind, however,
that in the experimental systems the elastic properties are likely
to be determined by the local organization of the molecules rather
than by the thickness variations. The discrete stretching energy is
\cite{parrinello1981polymorphic} $E_{s}^{dis.}=\sum_{T}\frac{YhA_{T}}{8\left(1+\nu\right)}\left(\frac{\nu}{1-\nu}\left(\mathrm{Tr}\hat{F}\right)^{2}+\mathrm{Tr}\hat{F}^{2}\right)$,
where $\hat{F}=\hat{\overline{g}}^{-1}\hat{g}-\hat{I}$ is the Cauchy-Green
strain tensor, $A_{T}$ is triangle area and the sum is carried out
over all triangles. Matrices $\hat{\overline{g}}$ and $\hat{g}$
are the discrete versions of the reference and actual metric tensors,
respectively, whose elements are the scalar products of the two vectors
spanning each triangle before and after the deformation. The bending
energy of the discrete mesh is \cite{Seung88} $\tilde{\kappa}\sum_{T_{i},T_{j}}\left(1-\mathbf{n}_{i}\cdot\mathbf{n}_{j}\right)$
with $\tilde{\kappa}=\frac{2}{\sqrt{3}}\kappa$ \cite{Seung88,schmidt2012}
and the sum runs over all pairs of the nearest neighbor triangles.
We point out that the last expression is actually a discrete version
of the continuum energy $\int dA\kappa\left(2H^{2}-K\right)$ \cite{Seung88}.
Therefore, we also need a discrete expression for the $\kappa\nu\int dAK$
term. Gaussian curvature at a vertex $i$ is \cite{Hu2011} $K_{i}\equiv\int_{A_{i}}KdA_{i}=2\pi-\sum_{T}\phi_{T}$,
where $\phi_{T}$ is the angle of the adjacent triangle $T$ at $i$
and $A_{i}=\frac{1}{3}\sum_{T}A_{T}$ is the associated vertex area.
Finally, \cite{Hu2011} $\kappa\nu\int dAK\to\kappa\nu\sum_{T}\frac{A_{T}}{3}\sum_{i}\frac{K_{i}}{A_{i}}$,
where the \emph{i}-sum runs over the three vertices of triangle $T$.

The surface mesh was constructed by building a triangulation with
$N_{v}$ vertices randomly (but evenly) distributed on a sphere. The
mesh with $N_{t}$ triangles was generated with the 3D Surface Mesh
Generation package in the CGAL library \cite{cgal}. In order to ensure
that the results are insensitive to the discretization details we
have performed independent runs starting from different initial configurations
with $2\times10^{3}$ to $5\times10^{3}$ vertices and $4\times10^{3}$
to $10^{4}$ triangles. Furthermore, we have validated our results
using regular triangulations of comparable sizes constructed according
to the prescription introduced by Caspar and Klug \cite{Caspar62}.
$N_{t}^{soft}=fN_{t}$, where $f=0.01\dots0.4$ is the fraction of
randomly chosen triangles were designated as soft. For a given random
mesh we have explored at least two different initial random distributions
of the soft triangles, removing a possible bias caused by a particular
choice of the initial distribution of the components. The reference
metric $\hat{\overline{g}}$ is an input parameter in our model and
was chosen to be that of the initial triangulation, which removes
the instability toward buckling into an icosahedron discussed in Ref.~
\cite{Lidmar03}. However, some residual stress is still present as
the stretching energy is minimized in the spherical configuration
while the bending energy favors flat faces. In order for the bending
energy to win, the vesicle wall has to be sufficiently thick. From
Eqs.~(\ref{eq:stretch}) and (\ref{eq:bend}) bending and stretching
energies are comparable at $h\approx a$, where $a$ is the unit length
set by the average edge length of a triangle. Our model implicitly
assumes that each triangle contains a sufficient number of microscopic
degrees of freedom such that the molecular details are of no importance.
In an typical amphiphilic system $a\approx5\mathrm{nm}$ sets the
length scale down to which the continuum description is applicable.
We set $h_{hard}=a$ ($a\ll R$ and the thin plate approximation is
valid) and note that in the $h_{hard}\to0$ limit the vesicle would
remain spherical regardless of the thickness ratio $\eta=\frac{h_{hard}}{h_{soft}}$.
Therefore, in our model a finite thickness is essential for faceting
to occur. Setting $\hat{\overline{g}}$ to be flat leads to interesting
effects which will be addressed elsewhere. In order to find the optimal
vesicle shape and the distribution of components simulated annealing
MC simulations were performed. A MC step consisted of two moves: (i)
A random displacement of a vertex was attempted, followed by (ii)
an attempted swap of the component types of a randomly chosen pair
of triangles. Both moves were accepted according to the Metropolis
algorithm. The system was heated up and cooled down using a linear
cooling protocol followed by three consecutive exponential cooling
runs. For each run, the configuration found to have the lowest energy
was recorded. A typical run consisted of $4\times10^{5}$ MC sweeps.
Note that the triangle type swap move is just a tool optimize the
component distribution and does not correspond to the actual redistribution
of the molecules that occurs during the vesicle assembly. The actual
kinetics of the self-assembly process is complex and beyond the scope
of this work.

\begin{figure}
\begin{centering}
\includegraphics[width=1\columnwidth]{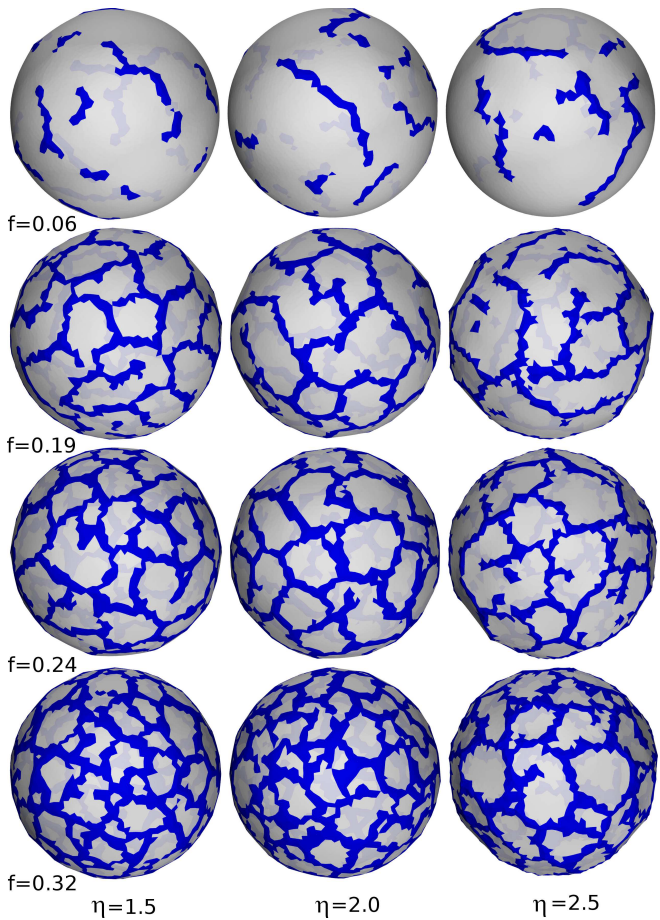}
\par\end{centering}

\caption{(color online) Snapshots of typical vesicle shapes as a function of
the thickness ratio $\eta=h_{hard}/h_{soft}=1.5$, $2.0$, $2.5$
and the fraction of the soft component $f=0.06$, $0.19$, $0.24$
and $0.32$. The soft component is shown in blue (dark) while the
hard component is white (light) and slightly transparent. For clarity,
width of the soft-component region has been enhanced. \label{fig:Optimal-shapes}}

\end{figure}

We set Poisson's to $\nu=\frac{1}{3}$, measure energies in units
of $Yh/a^{2}$ and measure mean and Gaussian curvatures in units of
$1/a$ and $1/a^{2}$, respectively. Note that the energy landscape
is complex and the simulated annealing is unlikely to find actual
ground state configurations. However, all obtained structures are
qualitatively reproducible and are thus referred to as \emph{typical}.
In Fig. \ref{fig:Optimal-shapes} we present snapshots of typical
vesicle shapes for a range of fractions $f$ of the soft component
and relative thicknesses, $\eta=\frac{h_{hard}}{h_{soft}}$. For $f\lesssim0.1$,
the soft component forms elongated, mutually disconnected ridges on
the vesicle surface. The vesicle remains nearly spherical with slight
distortions near the ridges. As $f$ increases to $\approx0.2$, ridges
begin to merge and facets develop. The onset of faceting is not sharp
and appears to be sensitive to $\eta$. If the amount of the soft
component is further increased the number of facets increases and
they become smaller in size. We speculate that for $f\gtrsim0.5$
this would no longer be the case and the soft component would form
islands; however, we have not explored this regime as it is not applicable
to the experimental systems of present interest. The hard-component
facets are, however, not perfectly flat and their curvature depends
on $\eta$. For $\eta\lesssim1.5$ the vesicle appears almost spherical,
while for $\eta\gtrsim2.5$ its shape is clearly polyhedral. 

\begin{figure}
\begin{centering}
\includegraphics[width=1\columnwidth]{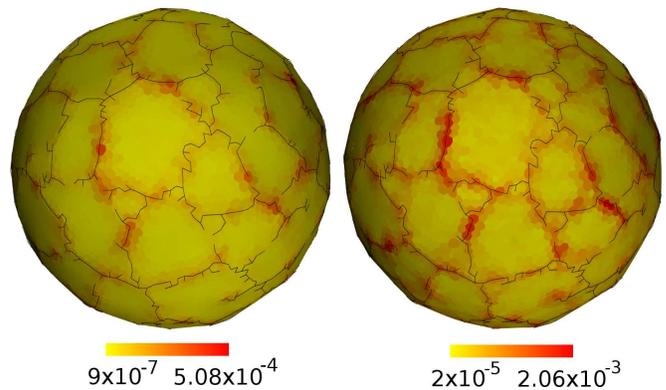}
\par\end{centering}

\caption{(color online) Distribution of stretching $E_{s}$ (left) and bending
$E_{b}$ (right) energies for $\eta=2.0$ and $f=0.25$. Black lines
indicate the soft boundaries between facets. For visualization purposes,
a lattice dual to the actual triangulation is shown. \label{fig:Distribution-of-energies}}

\end{figure}

In Fig. \ref{fig:Distribution-of-energies} we show the distribution
of the stretching and bending energies on the vesicle surface for
$\eta=2.0$ and $f=0.25$. As expected, the stretching energy is highest
along the boundaries where the material is soft. Note that the stretching
energy is particularly large near the kinks in the soft boundary suggesting
that such structures are probably not energetically preferred. Relaxing
these kinks requires a coherent global redistribution of the components
that is very hard to achieve in most numerical optimization schemes. 

\begin{figure}
\begin{centering}
\includegraphics[angle=270,width=1\columnwidth]{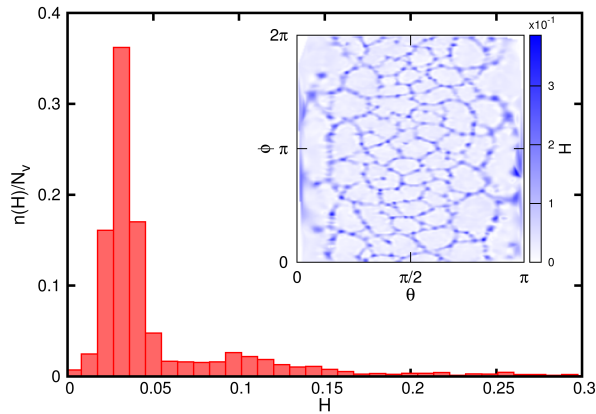}
\par\end{centering}

\caption{(color online) Distribution of the mean curvature, $H$, for $f=0.25$
and $\eta=2.0$ with two peaks corresponding to the hard facets (lower
$H$) and soft boundaries (higher $H$). \emph{Inset:} Projection,
$\phi\theta$-plot of $H$. Darker colors indicate high curvature
regions. Blurry distortions near the plot edges are an artifact of
the interpolation method. \label{fig:Mean-curvature}}

\end{figure}

\begin{figure}
\begin{centering}
\includegraphics[angle=270,width=1\columnwidth]{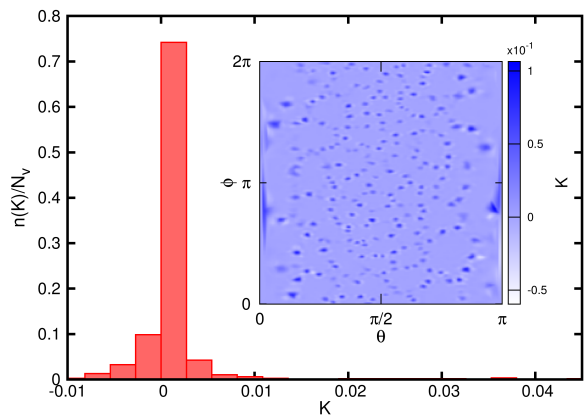}
\par\end{centering}

\caption{(color online) Distribution of the Gaussian curvature, $K$, for $f=0.25$
and $\eta=2.0$. \emph{Inset:} Projection, $\phi\theta$-plot of $K$.
Darker colors indicate high Gaussian curvature regions. Blurry distortions
near the plot edges are an artifact of the interpolation method. \label{fig:Gaussian-curvature}}

\end{figure}

In Fig. \ref{fig:Mean-curvature} we show the distribution of the
mean curvature for a vesicle with $f=0.25$ and $\eta=2.0$ with two
distinct peaks. The sharper peak at low $H$ corresponds to the facets,
while the weaker broader peak at a larger $H$ represents the curvature
at boundaries. A projection plot of $H$ clearly shows that the high
curvature is condensed along lines. Distribution of the Gaussian curvature
is shown in Fig. \ref{fig:Gaussian-curvature} with a sharp peak at
zero. Although $K=0$ on the facets, $H>0$ indicating that the facets
are not flat but locally cylindrical. Gaussian curvature is localized
along the soft grain boundaries as can be seen in the projection plot
in the inset in Fig. \ref{fig:Gaussian-curvature}. We note that a
number of vertices have $K<0$. Such regions locally resemble saddles
and are located along the soft-component ridges. A detailed analysis
of this effect would require a finer mesh and a more accurate minimization
technique than used here. 

We have implemented a general non-linear elastic model to analyze
faceting of solid-wall vesicles with soft domain boundaries into polyhedra
other than icosahedra. By choosing a suitable reference metric state,
we have removed the instability toward buckling into the icosahedron
seeded at the twelve five-fold defects \cite{Lidmar03}. While topological
defects are still present as required by the spherical topology their
effects are treated implicitly by assuming that they collect at the
boundary between facets making the vesicle wall locally soft. Consequently
the long-range effect of the stress produced by the defects are substantially
suppressed resulting in faceting into irregular polyhedra. We believe
that our model qualitatively explains the faceting observed in small
DPPC vesicles \cite{blaurock1979small} and vesicles coassembled by
oppositely charged amphiphiles \cite{Greenfield2009}. We conclude
by noting that other nonspherical shapes have also been observed in
multicomponent liquid vesicles \cite{Hu2011} and in vesicles assembled
of smectic polymer molecules \cite{jia2011smectic,Xing2011}. 

We thank C. Funkhouser, V. Jadhao, C. Thomas, and G. Vernizzi for
numerous comment on the manuscript. Numerical simulations were performed
using the Northwestern High Performance Computing Cluster - Quest.
This works was supported by the US Department of Energy Award DEFG02-08ER46539
and by the US Air Force Office of Research NSSEFF Award.

\bibliographystyle{apsrev}

\end{document}